\documentclass[10pt]{article}
\usepackage{amsmath}
\usepackage{amssymb}
\usepackage{epsfig}
\usepackage{euscript}
\usepackage{fancybox}
\usepackage{color}

\def\mathswitchr#1{\relax\ifmmode{\mathrm{#1}}\else$\mathrm{#1}$\fi}

\def\rQCED{{\rm QCED}}

\newcommand {\pslash}{\hbox{$\not\hbox{\kern-2.3pt $p$}$}}

\usepackage{cite}

\usepackage{epic}

\def\alf1{ {\alpha\over\pi} }

\begin{document}
\begin{titlepage}
\begin{flushright}
{\bf BU-HEPP-08-11}\\
{\bf Aug., 2008}\\
\end{flushright}
 
\begin{center}
{\Large Precision QED$\otimes$QCD Resummation Theory for LHC Physics: Status and Update$^{\dagger}$}
\end{center}

\vspace{2mm}
\begin{center}
{\bf   B.F.L. Ward$^a$, S. Joseph$^a$, S. Majhi$^a$ and S.A. Yost$^b$}\\
\vspace{2mm}
{\em $^a$Department of Physics,\\
 Baylor University, Waco, Texas, USA}\\
{\em $^b$Department of Physics,\\
  The Citadel, Charleston, South Carolina, USA}\\
\end{center}

\vspace{5mm}
\begin{center}
{\bf   Abstract}
\end{center}
We present the elements of the IR-improved DGLAP-CS theory as it relates to 
the new MC friendly exponentiated scheme for precision calculation of higher 
order corrections to LHC physics in which IR singularities from both QED and 
QCD are canceled to all orders in $\alpha$ and in $\alpha_s$ 
simultaneously in the 
presence of rigorous shower/ME matching. We present the first MC data 
comparing the implied new showers themselves with the standard ones using 
the HERWIG6.5 MC event generator as a test case at LHC energies. As expected, 
the IR-improved shower re-populates lower values of the energy fraction z and 
lower values of the attendant $p_T$ for 
the standard HERWIG6.5 input parameters. 
Possible phenomenological implications are discussed.
\\
\vskip 20mm
\vspace{10mm}
\renewcommand{\baselinestretch}{0.1}
\footnoterule
\noindent
{\footnotesize
\begin{itemize}
\item[${\dagger}$]
Work partly supported by US DOE grant DE-FG02-05ER41399 and 
by NATO Grant PST.CLG.980342.
\end{itemize}
}

\end{titlepage}

\def\Kmax{K_{\rm max}}\def\ieps{{i\epsilon}}\def\rQCD{{\rm QCD}}
\renewcommand{\theequation}{\arabic{equation}}
\font\fortssbx=cmssbx10 scaled \magstep2
\renewcommand\thepage{}
\parskip.1truein\parindent=20pt\pagenumbering{arabic}\par
\section{\bf Introduction}\label{intro}\par
With the advent of the LHC, we enter the era of precision QCD, by which we mean
predictions for QCD processes at the total theoretical precision 
tag of $1\%$ or better. 
The attendant requirement for this theoretical precision is 
control of the ${\cal O}(\alpha_s^2L^{n_1},\alpha_s\alpha L^{n_2},\alpha^2L^{n_3})$, ~$n_1 = 0, 1, 2, ~n_2 = 1, 2, ~n_3 = 2$ corrections 
in the presence of realistic
parton showers, on an event-by-event basis -- here, $L$ is a generic big log. This is the objective
of our approach to precision QCD theory, 
which for example will be needed for the expected ~2\% experimental precision
~\cite{lhclum1} at the LHC for processes such as $pp\rightarrow V+m(\gamma)+n(G)+X\rightarrow \bar\ell\ell'+m'(\gamma)+n(G)+X$,~
$V=W^\pm,Z$, and $\ell=e,\mu, ~\ell'=\nu_e,~\nu_\mu(e,\mu)$ for $V=W^+(Z)$ respectively,
and $\ell=\nu_e,~\nu_\mu,~\ell'=e,\mu$ respectively for $V=W^-$.
Here, we present the elements
of our approach and its recent applications in MC event generator studies,
which are still preliminary.\par
At such a precision as we have as our goal, issues such as the role of QED are an integral part of the discussion and we deal with this by the simultaneous
resummation of QED and QCD large infrared effects, $QED\otimes QCD$ resummation
~\cite{qced} in the presence of parton showers, to be realized on an 
event-by-event basis by Monte Carlo methods. This is reviewed
in the next Section. Let us note already that in Refs.~\cite{qedeffects}
it has been shown that QED evolution enters at the $\sim 0.3\%$ level
for parton distributions and that in Refs.~\cite{radcor-ew} it has been
shown that EW (large Sudakov logs, etc.) effects at LHC energies, 
as W's and Z's are almost
massless on the TeV scale, can enter at the several \% level -- 
such corrections
must treated systematically before any claim of 1\% preciaion
can be taken seriously. We are presenting a framework in which
this can be done. The new amplitude-based resummation
algebra then leads to a new scheme for calculating 
hard hadron-hadron scattering
processes, IR-improved DGLAP-CS theory~\cite{irdglap1} 
for parton distributions, kernels,
reduced cross sections with the appropriate shower/ME matching.
This is summarized in Section III. In this latter Section, 
with an eye toward technical precision cross checks plus 
possible physical effects of heavy quark masses, we also
deal with  the issue of quark masses as collinear 
regulators~\cite{sac-no-go,cat1,qmass} as an alternative~\cite{qmass-bw}
to the usual practice of setting all initial state quark masses to zero in
calculating ISR (initial state radiation) effects in 
higher order QCD corrections. We also discuss in Section III 
the relationship between our resummation algebra and that
of Refs.~\cite{cattrent,scet}, as again such comparisons will be necessary
in assessing the ultimate theoretical precision tag. 
In Section IV, we illustrate
recent results we have obtained for the effects of our new approach
on the parton showers as they are generated with the 
HERWIG6.5 MC~\cite{herwig}. Extensions of such studies to 
PYTHIA~\cite{pythia} and MC@NLO~\cite{mcnlo} are in progress.
Section V contains summary remarks.\par
As a point of reference, in Ref.~\cite{scott1} it has been argued that
the current state-of-the-art theoretical precision tag on single Z
production at the LHC is $(4.1\pm0.3)\%=(1.51\pm 0.75)\%(QCD)\oplus 3.79(PDF)\oplus 0.38\pm 0.26(EW)\%$,
where the results of Refs.~\cite{cteq,mrst,mcnlo,fewz,resbos,horace,photos} have been
used in this precision tag determination.\footnote{Recently, the 
analogous estimate for single W production has been given in Ref.~\cite{scott2} -- it is $\sim 5.7$\%.}\par
\section{QED$\otimes$QCD Resummation}
In refs.~\cite{qced}, we have extended the YFS theory to the 
simultaneous exponentiation of the large 
IR terms in QCD and the exact IR divergent terms in QED, so that
for the prototypical subprocesses
$\bar{Q}'Q\rightarrow \bar{Q}'''Q''+m(G)+n(\gamma)$
we arrive at the new result
{\small
\begin{equation}
\begin{split}
d\hat\sigma_{\rm exp} &= e^{\rm SUM_{IR}(QCED)}\\
   &\sum_{{m,n}=0}^\infty\frac{1}{m!n!}\int\prod_{j_1=1}^m\frac{d^3k_{j_1}}{k_{j_1}} 
\prod_{j_2=1}^n\frac{d^3{k'}_{j_2}}{{k'}_{j_2}}
\int\frac{d^4y}{(2\pi)^4}\\&e^{iy\cdot(p_1+q_1-p_2-q_2-\sum k_{j_1}-\sum {k'}_{j_2})+
D_\rQCED} \\
&\tilde{\bar\beta}_{m,n}(k_1,\ldots,k_m;k'_1,\ldots,k'_n)\frac{d^3p_2}{p_2^{\,0}}\frac{d^3q_2}{q_2^{\,0}},
\end{split}
\label{qced1}
\end{equation}}\noindent
where the new YFS~\cite{yfs,yfs1} residuals, defined in Ref.~\cite{qced}, 
$\tilde{\bar\beta}_{m,n}(k_1,\ldots,k_m;k'_1,\ldots,k'_n)$, with $m$ hard gluons and $n$ hard photons,
represent the successive application
of the YFS expansion first for QCD and subsequently for QED. The
functions ${\rm SUM_{IR}(QCED)},D_\rQCED$ are determined
from their analogs ${\rm SUM_{IR}(QCD)},D_\rQCD$ in Ref.~\cite{qcdref}
via the substitutions
{\small
\begin{eqnarray}
B^{nls}_{QCD} \rightarrow B^{nls}_{QCD}+B^{nls}_{QED}\equiv B^{nls}_{QCED},\cr
{\tilde B}^{nls}_{QCD}\rightarrow {\tilde B}^{nls}_{QCD}+{\tilde B}^{nls}_{QED}\equiv {\tilde B}^{nls}_{QCED}, \cr
{\tilde S}^{nls}_{QCD}\rightarrow {\tilde S}^{nls}_{QCD}+{\tilde S}^{nls}_{QED}\equiv {\tilde S}^{nls}_{QCED}
\label{irsub}
\end{eqnarray}}  
everywhere in expressions for the
latter functions given in Refs.~\cite{qcdref} -- see Ref.~\cite{qced} for
the details of this substitution. It can be readily established~\cite{qced}
that the QCD dominant corrections happen an order of magnitude earlier in time
compared to those of QED so that the leading term $\tilde{\bar\beta}_{0,0}$ 
already gives us a good estimate of the size of the effects we study.\par
Important in any total theoretical precision is knowledge of possible
systematic issues associated with ones methods. This entails the
relationship between different approaches to the same classes of corrections
and moves us to the relationship between our approach to QCD resummation
and the more familiar approach in Refs.~\cite{cattrent}. It has been shown
in Ref.~\cite{geor1} that the latter approach is entirely equivalent to
the approach in Refs.~\cite{scet}. Establishing the relationship between our approach and that in Refs.~\cite{cattrent} will then suffice to relate all three approaches.\par
In Ref.~\cite{madg} 
the more familiar resummation for soft gluons in Refs.~\cite{cattrent}
is applied to a general $2\rightarrow n$ parton
process [f] at hard scale Q,
$f_1(p_1,r_1)+f_2(p_2,r_2)\rightarrow f_3(p_3,r_3)+f_4(p_4,r_4)+\cdots+f_{n+2}(p_{n+2},r_{n+2})$, where the $p_i,r_i$ label 4-momenta and color indices respectively, with all parton masses set to zero to get
\begin{equation}
\begin{split}
{\cal M}^{[f]}_{\{r_i\}}&=\sum^{C}_{L}{\cal M}^{[f]}_L(c_L)_{\{r_i\}}\\
&= J^{[f]}\sum^{C}_{L}S_{LI}H^{[f]}_I(c_L)_{\{r_i\}},
\end{split}
\label{madg1}
\end{equation}
where repeated indices are summed, 
$J^{[f]}$ is the jet function,
$S_{LI}$ is the soft function which describes
the exchange of soft gluons between the external lines, and
$H^{[f]}_I $ is the hard coefficient function.
The attendant infrared and collinear 
poles are calculated to 2-loop order.
To make contact with our approach,
identify in 
$\bar{Q}'Q\rightarrow \bar{Q}'''Q''+m(G)$ in (\ref{qced1}) $f_1=Q,\bar{Q}', f_2=\bar{Q}',
f_3=Q'', f_4=\bar{Q}''', \{f_5,\cdots,f_{n+2}\}=\{G_1,\cdots,G_m\}$ so that
$n=m+2$ here.
Observe the following:
\begin{itemize}
\item By its definition in eq.(2.23) of Ref.~\cite{madg}, 
the anomalous dimension
of the matrix $S_{LI}$ does not contain any of the diagonal effects described by our infrared functions $\Sigma_{IR}(QCD)$ and $D_{QCD}$.
\item By its definition in eqs.(2.5) and (2.7) of Ref.~\cite{madg}, the jet function $J^{[f]}$ contains the exponential of the virtual infrared function $\alpha_s\Re{B}_{QCD}$, so that we have to take care that we do not double count when we
use (\ref{madg1}) in (\ref{qced1}) and the equations that lead thereto.
\end{itemize} 
It follows that, referring to our analysis in Ref.~\cite{annphys08},
we identify $\bar\rho^{(m)}$ in eq.(73) in this latter reference
in our theory as{\small 
\begin{equation}
\begin{split}&\bar\rho^{(m)}(p_1,q_1,p_2,q_2,k_1,\cdots,k_m)
=\overline\sum_{colors,spin}|{\cal M}^{'[f]}_{\{r_i\}}|^2\\
&\equiv \sum_{spins,\{r_i\},\{r'_i\}}\mathfrak{h}^{cs}_{\{r_i\}\{r'_i\}}|\bar{J}^{[f]}|^2\sum^{C}_{L=1}\sum^{C}_{L'=1}S^{[f]}_{LI}H^{[f]}_I(c_L)_{\{r_i\}}\left(S^{[f]}_{L'I'}H^{[f]}_{I'}(c_{L'})_{\{r'_i\}}\right)^\dagger,
\end{split}
\label{madg2}
\end{equation}
where here we defined $\bar{J}^{[f]}=e^{-\alpha_s\Re{B}_{QCD}}J^{[f]}$, 
and we introduced the color-spin density matrix for the initial state, $\mathfrak{h}^{cs}$.
Here, we recall (see Refs.~\cite{irdglap1,annphys08}, for example)
that in our theory, we have
\begin{eqnarray}
  d\hat\sigma^n = \frac{e^{2\alpha_sReB_{QCD}}}{n !}\int\prod_{m=1}^n
\frac{d^3k_m}{(k_m^2+\lambda^2)^{1/2}}\delta(p_1+q_1-p_2-q_2-\sum_{i=1}^nk_i)
\nonumber\\       
\bar\rho^{(n)}(p_1,q_1,p_2,q_2,k_1,\cdots,k_n)
\frac{d^3p_2d^3q_2}{p^0_2 q^0_2},
\label{diff1}
\end{eqnarray}}
for n-gluon emission. It follows that
we can repeat thus our usual steps (see Ref.~\cite{irdglap1,annphys08})
to get the QCD corrections in our formula
(\ref{qced1}), without any double counting of effects. This use of the
results in Ref.~\cite{madg} is in
progress.\par
\section{IR-Improved DGLAP-CS Theory: Applications}
In Refs.~\cite{irdglap1,annphys08} it has been shown that application of
the result (\ref{qced1}) to all aspects of the standard formula
for hard hadron-hadron scattering processes, 
\begin{equation}
\sigma =\sum_{i,j}\int dx_1dx_2F_i(x_1)F_j(x_2)\hat\sigma(x_1x_2s)
\label{sigtota}
\end{equation}
where we the $\{F_i(x)\}$ and $\hat\sigma$ 
denote the parton densities and reduced
cross section respectively, leads one to its application
to the DGLAP-CS theory itself for the kernels
which govern the evolution of the parton densities in addition to the
the implied application to the respective 
hard scattering reduced cross section. The result is a new set of
IR-improved kernels~\cite{irdglap1},
\begin{align}
P_{qq}(z)&= C_F F_{YFS}(\gamma_q)e^{\frac{1}{2}\delta_q}\left[\frac{1+z^2}{1-z}(1-z)^{\gamma_q} -f_q(\gamma_q)\delta(1-z)\right],\\
P_{Gq}(z)&= C_F F_{YFS}(\gamma_q)e^{\frac{1}{2}\delta_q}\frac{1+(1-z)^2}{z} z^{\gamma_q},\\
P_{GG}(z)&= 2C_G F_{YFS}(\gamma_G)e^{\frac{1}{2}\delta_G}\{ \frac{1-z}{z}z^{\gamma_G}+\frac{z}{1-z}(1-z)^{\gamma_G}\nonumber\\
&\qquad +\frac{1}{2}(z^{1+\gamma_G}(1-z)+z(1-z)^{1+\gamma_G}) - f_G(\gamma_G) \delta(1-z)\},\\
P_{qG}(z)&= F_{YFS}(\gamma_G)e^{\frac{1}{2}\delta_G}\frac{1}{2}\{ z^2(1-z)^{\gamma_G}+(1-z)^2z^{\gamma_G}\}.
\label{dglap19}
\end{align}
in the standard notation, where  
\begin{align}
\gamma_q &= C_F\frac{\alpha_s}{\pi}t=\frac{4C_F}{\beta_0}\\
\delta_q&=\frac{\gamma_q}{2}+\frac{\alpha_sC_F}{\pi}(\frac{\pi^2}{3}-\frac{1}{2})\\
\gamma_G &= C_G\frac{\alpha_s}{\pi}t=\frac{4C_G}{\beta_0}\\
\delta_G&=\frac{\gamma_G}{2}+\frac{\alpha_sC_G}{\pi}(\frac{\pi^2}{3}-\frac{1}{2})
\label{dglap9}
\end{align}
and 
\begin{equation}
F_{YFS}(\gamma_q)=\frac{e^{-C_E\gamma_q}}{\Gamma(1+\gamma_q)},
\label{dglap10}
\end{equation}
so that
\begin{align}
f_q(\gamma_q)&=\frac{2}{\gamma_q}-\frac{2}{\gamma_q+1}+\frac{1}{\gamma_q+2}\\
f_G(\gamma_G)&=\frac{n_f}{C_G}\frac{1}{(1+\gamma_G)(2+\gamma_G)(3+\gamma_G)}+
\frac{2}{\gamma_G(1+\gamma_G)(2+\gamma_G)}\\
&\qquad +\frac{1}{(1+\gamma_G)(2+\gamma_G)}
+\frac{1}{2(3+\gamma_G)(4+\gamma_G)}\\
&\qquad +\frac{1}{(2+\gamma_G)(3+\gamma_G)(4+\gamma_G)}.
\label{dglap19a}
\end{align}
Here,
$C_E = 0.5772...$ is Euler's constant and $\Gamma(w)$ is the Euler Gamma
function.
We see that the kernels are integrable at the IR endpoints and this admits
a more friendly MC implementation, which is in progress.\par
Some observations are in order. First,
We note that the connection of (\ref{dglap19}) with the
higher order kernel results in Refs.~\cite{high-ord-krnls} is immediate
and has been shown in Refs.~\cite{irdglap1,annphys08}. Second, there is no
contradiction with the standard Wilson expansion, as the terms we resum
are not in that expansion by its usual definition. Third, we do not
change the predicted cross section: we have a new scheme such that the
cross section in (\ref{sigtota}) becomes
\begin{equation}
\sigma  =\sum_{i,j}\int dx_1dx_2{F'}_i(x_1){F'}_j(x_2)\hat\sigma'(x_1x_2s)
\label{sigtotb}
\end{equation} 
order by order in perturbation theory, where 
$\{P^{exp}\}$ factorize $\hat\sigma_{\text{unfactorized}}$ to yield
$\hat\sigma'$ and its attendant parton densities $\{{F'}_i\}$. 
Fourth, when one solves for the effects of the exponentiation 
in (\ref{dglap19}) on the actual evolution of the parton densities
from the typical reference scale of $Q_0\sim 2$GeV to $Q=100$ GeV one 
finds~\cite{irdglap1,annphys08} shifts of $\sim 5\%$ for the NS n=2 moment
for example, which is thus of some phenomenological interest-- see for example
Ref.~\cite{carli}. 
Finally, we note that we have used~\cite{qced} the result
(\ref{qced1}) for single Z production with leptonic decay at the
LHC (and at FNAL) to focus on the ISR alone, for definiteness 
and we find agreement with the literature in  
Refs.~\cite{baurall,ditt,zyk} for exact ${\cal O}(\alpha)$ results and 
Refs.~\cite{van1,van2,anas} for exact ${\cal O}(\alpha_s^2)$ results,
with a threshold QED effect of ~0.3\%, similar to that found
for the parton evolution itself from QED in Refs.~\cite{qedeffects}.
Evidently, any 1\% percision tag must account for all such effects.\par
\subsection{Shower/ME Matching}
In using (\ref{qced1}) in (\ref{sigtotb}) for $\hat\sigma'(x_ix_j)$, we
intend to combine our exact extended 
YFS calculus with HERWIG~\cite{herwig} and 
PYTHIA~\cite{pythia} as follows: they generate a parton shower starting from $(x_1,x_2)$ at the factorization scale $\mu$ after this point is provided by the
$\{F'_i\}$ and we may use~\cite{qced} either a $p_T$-matching scheme
or a shower-subtracted residual scheme where the respective
new residuals $\{\hat{\tilde{\bar\beta}}_{n,m}(k_1,\ldots,k_n;k'_1,\ldots,k'_m)\}$ are obtained by expanding the shower formula and the result in (\ref{qced1})
on product and requiring the agreement with exact results to the specified order.\footnote{ See Ref.~\cite{baloss} for a realization of the shower subtracted residual scheme in the context of QED parton showers.} This combination of theoretical constructs can be 
systematically improved with
exact results order-by-order in $\alpha_s,\alpha$, with exact phase space.
\footnote{The current state of the art for such shower/ME matching is
given in Refs.~\cite{mcnlo}, which realizes exactness at
${\cal O}(\alpha_s)$.}
The recently developed new parton evolution algorithms in Refs.~\cite{jad-skrz}
may also be used here.\par
The issue of the non-zero quark masses in the initial state radiation is
present when one wants 1\% precision, as we know that 
the parton densities for the heavy quarks are all different
and the generic size of mass corrections for bremsstrahlung is $\alpha_s/\pi$
for cross sections~\cite{lee-naun}, so that one would like to
know whether regularizing a zero-mass ISR radiation result with
dimensional methods, carrying through the factorization procedure gives the
same result as doing the same calculation with the physical, non-zero mass
of the quark and again carrying through the factorization procedure to
the accuracy $\alpha_s^2/\pi^2$, for example. Until the analysis in Ref.~\cite{qmass-bw}, this cross check was not possible because 
in Refs.~\cite{sac-no-go,cat1} it was
shown that there is a lack of Bloch-Nordsieck cancellation in the ISR
at ${\cal O}(\alpha_s^2)$ unless the radiating quarks are massless.
The QCD resummation algebra, as used in (\ref{qced1}), 
allows us to obviate~\cite{qmass-bw} this theorem,
so that now such cross checks are possible and they are in progress.\par
\subsection{Sample MC data: IR-Improved Kernels in HERWIG6.5}  
We have preliminary results on IR-improved showers in HERWIG6.5:
we compare the z-distributions and the $p_T$ of the IR-improved and usual
DGLAP-CS showers in the Figs.~1,~2,~3. As we would expect, the IR-improved shower re-populates the soft region in both variables. 
The details of the implementation procedure and the respective new version
of HERWIG6.5, HERWIG6.5-YFS, will appear elsewhere~\cite{elsewh}.
The analogous implementations in PYTHIA and MC@NLO are in progress, as are
comparisons with IR safe observables.\par
\begin{figure}
\begin{center}
\epsfig{file=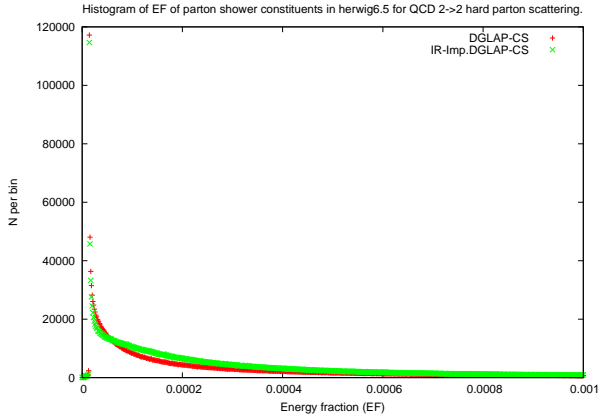,angle=270,width=80mm}
\end{center}
\caption{\baselineskip=7mm     The z-distribution shower comparison in HERWIG6.5 -- preliminary results.}
\label{fighw1}
\end{figure}
\begin{figure}
\begin{center}
\epsfig{file=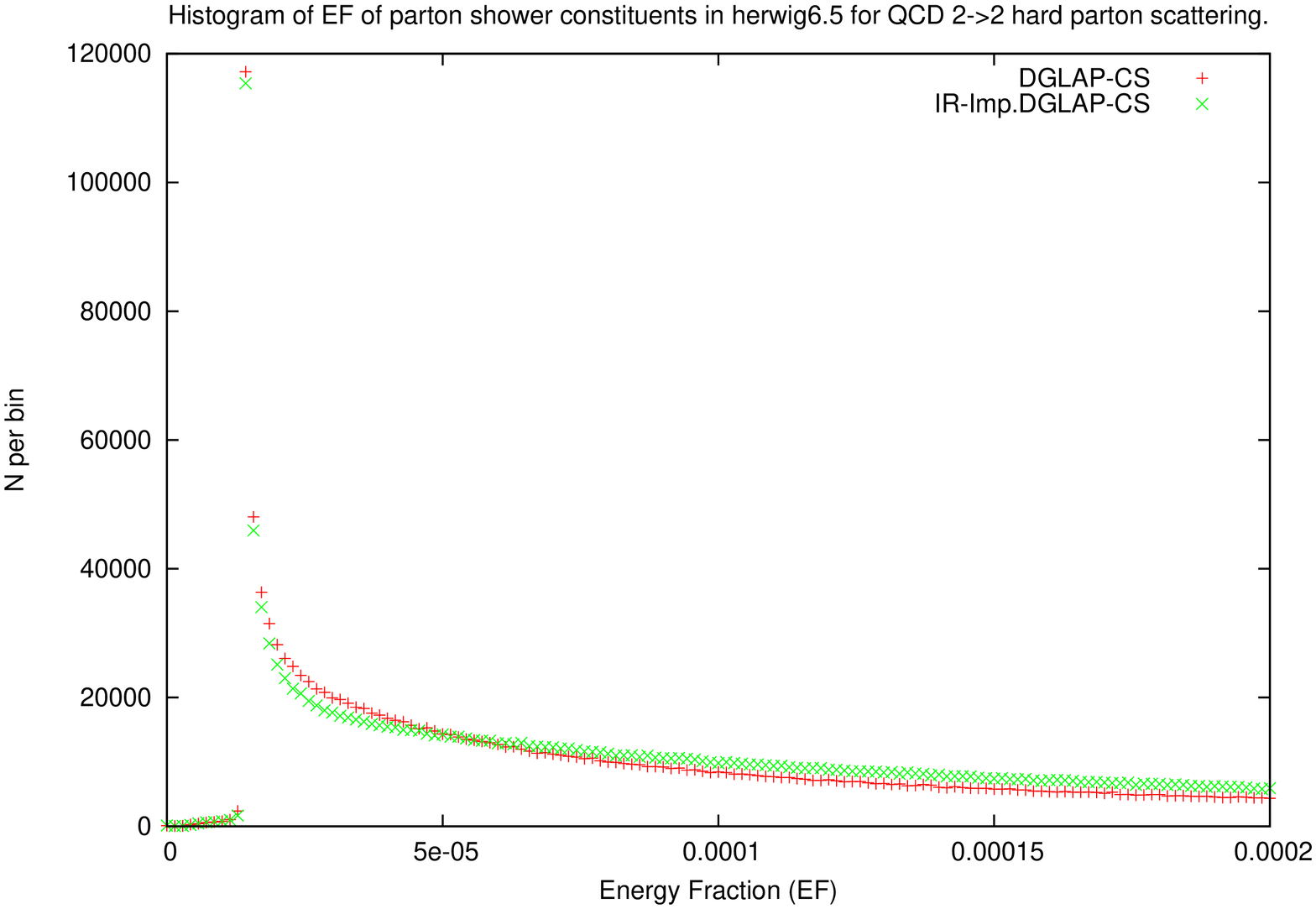,angle=270,width=80mm}
\end{center}
\caption{\baselineskip=7mm     The z-distribution shower comparison in HERWIG6.5 at small z -- preliminary results.}
\label{fighw2}
\end{figure}
\begin{figure}
\begin{center}
\epsfig{file=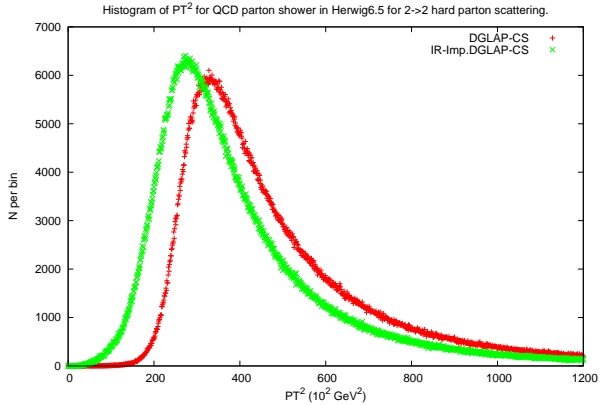,angle=270,width=80mm}
\end{center}
\caption{\baselineskip=7mm     The $p_T$-distribution shower comparison in HERWIG6.5 -- preliminary results.}
\label{fighw3}
\end{figure}
\section{Conclusions}
The theory of Ref.~\cite{yfs} extends to the joint resummation of QED and QCD
with proper shower/ME matching built-in. For the simultaneous QED$\otimes$QCD
resummed theory, full MC event generator realization is 
open: a firm basis for 
the complete ${\cal O}(\alpha_s^2,\alpha\alpha_s,\alpha^2)$ MC results
needed for precision LHC physics has been demonstrated and all the latter
are in progress -- see Refs.~\cite{hypgeo} for new results on $\epsilon$
expansions for the higher order Feynman integrals needed to isolate the
residuals in our approach for example. This allows cross check 
between residuals isolated with the quark masses as regulators, 
something now allowed by
the result in Ref.~\cite{qmass-bw}, and those isolated in 
dimensional regularization
for the massless quark limit. Such cross checks are relevant 
for precision
QCD theory.
The first MC data have been shown
with IR-improved showers in HERWIG6.5. The spectra are softer as expected. We look forward to the detailed comparison with IR safe observables 
as generated with IR-improved and with the usual showers -- this will appear elsewhere.~\cite{elsewh}. Already, semi-analytical results at the $\tilde\bar\beta_{0,0}^{0,0}$ are consistent with the literature on single Z production, while a cross check 
for the analogous W production is near. As the QED is at 0.3\% at threshold,
it is needed for 1\% precision.\par
\section*{Acknowledgments}
One of us (B.F.L.W) acknowledges helpful discussions with Prof. Bryan Webber
and Prof. M. Seymour. B.F.L. Ward also thanks Prof. L. Alvarez-Gaume and Prof. W. Hollik for the support and kind hospitality of the CERN TH Division and of MPI, Munich, respectively, while this work was in progress.

\end{document}